\documentclass[12pt, a4paper]{article}
\usepackage{url}
\usepackage{verbatim}

\title{Health Information Retrieval \\
\Large{State of the art report}
}

\author{Carla Teixeira Lopes\\
{\normalsize Doctoral Programme in Informatics Engineering} \\
{\normalsize Faculdade de Engenharia da Universidade do Porto, Portugal}\\
{\small carla.lopes@fe.up.pt}
\medskip \\
{\normalsize Supervisor: Maria Cristina Ribeiro}
\medskip \\
}

\date{ 
{\normalsize July 2008}
}

\begin{document}
\pagestyle{plain}
\pagenumbering{roman}
\maketitle

\begin{abstract}

This report provides an overview of the field of Information Retrieval (IR) in healthcare. It does not aim to introduce general concepts and theories of IR but to present and describe specific aspects of Health Information Retrieval (HIR). After a brief introduction to the more broader field of IR, the significance of HIR at current times is discussed. Specific characteristics of Health Information, its classification and the main existing representations for health concepts are described together with the main products and services in the area (e.g.: databases of health bibliographic content, health specific search engines and others). Recent research work is discussed and the most active researchers, projects and research groups are also presented. Main organizations and journals are also identified.

\end{abstract}

\pagebreak
\tableofcontents
\pagebreak

\cleardoublepage
\pagenumbering{arabic}

\section{Introduction}

This document describes the state of the art on Health Information Retrieval (HIR), a field of study where Information Retrieval (IR) techniques are applied in the specific domain of Health Information. It starts with a brief introduction to the broader field of IR (Section \ref{IR}) where its definition and a short description of its evolution through the years is given, followed by an introduction to the more specific field of HIR (Section \ref{HIR}). Section \ref{HI} presents a classification of Health Information in several categories where several examples of products and services are given. In Section \ref{HCR} the main existing representations for health concepts are described, from terminologies to ontologies. Section \ref{HSSE} enumerates and briefly describes the main health specific search engine and Section \ref{RW} describes some of the recent research work developed in the area. In sections \ref{People}, \ref{RG}, \ref{RP}, \ref{MO} and \ref{MJ} are presented major figures in the area along with its research groups, research projects, main organizations and also journals.


\subsection{Information Retrieval}
\label{IR}

\subsubsection{Definition}

Information retrieval (IR) is a broad and loosely-defined term \cite{Rijsbergen_book_1979, Manning_book_2008} that has become accepted with the work published by Cleverdon, Salton, Sparck Jones, Lancaster and others \cite{Rijsbergen_book_1979}. One of the earliest definitions, that dates from the 1960's, characterizes it as an area concerned with the structure, analysis, organization, storage, searching and retrieval of information  \cite{Salton_book_1968}. More recently, Baeza-Yates et al. \cite{Baeza-Yates_book_1999} presented a similar definition where IR deals with the representation, storage, organization and access to information items. An even more recent definition presents IR as ``finding material (usually documents) of an unstructured nature (usually text) that satisfies an information need from within large collections (usually stored on computers)'' \cite{Manning_book_2008}.

Being an imprecise distinction, it is frequent to distinguish IR from Data Retrieval. Rijsbergen \cite{Rijsbergen_book_1979} does it in eight different perspectives: matching, inference, model, classification, query language, query specification, items wanted and error response. Baeza-Yates at al \cite{Baeza-Yates_book_1999} also make this distinction defining data retrieval as retrieving all objects that satisfy clearly defined conditions where an erroneous object among the retrieved ones means a total failure. In IR it is not a major problem to identify non-relevant documents in the retrieved set of documents because this field of study deals with information that is not structured and that may be semantically ambiguous (the opposite of Data Retrieval) \cite{Baeza-Yates_book_1999, Manning_book_2008, Allan_SIGForum_2003}. Research in database systems is usually associated with the Data Retrieval field.


\subsubsection{Brief history}


The first manual IR systems appeared with the Sumerians, in the beginning of 3000 BC, when they constructed areas for storing and classifying written materials (cuneiform inscriptions - one of history's oldest writing system) to assist the operation of various social groups \cite{libraries_url_und}. As time went by, inventions like paper, the printed press raised the importance of systems to store, manage and retrieve information. After the invention of computers, in 1945, Vannevar Bush wrote the article ``As We May Think'' \cite{Bush_TheAtlantic_1945} in which he criticizes the artificiality of the epoch's indexing systems. He idealizes a system that operates through association (as the human mind) and is mechanized so it may be consulted with speed and flexibility. This was the first conception of an automatic IR system. The term Information Retrieval was coined later, in 1950, by Calvin Mooers \cite{mooers_AD_1950}.

Since the late 1950s, the field of IR has evolved through several relevant works such as: H.P. Luhn's work \cite{Luhn_JRD_1957}; the SMART system by Salton and his students \cite{Salton_und_1971} where some important IR concepts (such as the vector space model and relevance feedback) were developed; Cleverdon's Cranfield evaluation model \cite{Cleverdon_Aslib_1967}; Sparck  Jones’ development of idf \cite{Jones_JD_1972} and the probabilistic models by Robertson \cite{Robertson_JASIS_1976} and Croft \cite{croft_JD_1979, Turtle_TOIS_1991}. In 1992, with the beginning of the Text REtrieval Conference (TREC) \cite{trec_url_und}, that provide the necessary infrastructure for large-scale evaluation, allowed the modification of old models/techniques and the proposal of new ones. 
 
With the development of the World Wide Web (Web), the interest in the IR field has spread from information specialists (librarians, the legal community and other) \cite{Allan_SIGForum_2003,  callan_sigirforum_2007,  Manning_book_2008} to a broader audience \cite{Allan_SIGForum_2003}. The larger interest on the area together with the increase in the amount of information available and in users' requirements have contributed to significant progresses on the field that has evolved from a simple document retrieval to a broad range of related areas. Some of these areas are: question-aswering, cross-lingual search, topic detection and tracking, summarization, multimedia retrieval and others \cite{Allan_SIGForum_2003, callan_sigirforum_2007}.
 
Despite the recent and numerous developments, IR is far from being a ``solved problem" \cite{callan_sigirforum_2007}. On one side, information is being produced more than ever -- a report of a study developed in UC Berkeley's School of Information Management and Systems  \cite{Lyman_url_2003} estimates that the amount of new information stored on paper, film, magnetic, and optical media has doubled and that the information on the Web has tripled from 2000 to 2003. 
On the other side, the ways people produce, search, manage and use information is rapidly evolving.

\subsection{Health Information Retrieval}
\label{HIR}


Probably by the same reasons that made the general IR field evolve so much in the last years, research and interest on the application of this field's techniques to the specific domain of health and biomedicine have also grew. The Web and its applications have profoundly changed the availability and ease of access to health information, not only to health professionals but also to consumers. 

To health professionals, applications providing an easy access to validated and up-to-date health knowledge are of great importance to the dissemination of knowledge and have the potential to impact the quality of care provided by health professionals. On the other side, the Web opened doors to the access of health information by patients, their family and friends, making them more informed and changing their relation with health professionals. According to Adam Bosworth \cite{Bosworth_AMIA_2007}, in a good health system, consumers should be able to discover the most relevant information possible, should have access to health support personalized services and should be able to learn from and educate those in similar health circumstances.

To professionals, one of the main and oldest IR applications is PubMed from the US National Library of Medicine (NLM) that gives access to the world's medical research literature. To consumers, health information is available through different services and with different quality. Lately, the control over and access to health information by consumers has been a hot topic, with plenty government initiatives all over the world that aim to improve consumer health giving consumers more information and making easier the sharing of patient records. Some of the main IR related companies, like Google and Microsoft are beginning to invest in this area. Google launched, at May 2008, a service entitled Google Health \cite{googlehealth_url_und} and Microsoft has also launched, at October 2007, a service entitled Microsoft HealthVault \cite{healthvault_url_und}. Revolution Health \cite{revolutionhealth_url_und} is another service of this type provided recently from Steve Case's (AOL founder) company. Generally, the movement of a consumer-driven healthcare is about giving consumer more power to manage their health. This can be done by giving them access to their health record anywhere/anytime, making easy to find all needed information, providing online tools that enable personalized advice and other approaches. 

Currently, the Web is already used in large scale to access health information. A 2006 study \cite{Fox_und_2006} concluded that  ``eight in ten USA internet users go online for health information''. They also found that: health searches are as popular as reading blogs or using the Internet to look up a phone number or address;  ``the typical health information session starts at a search engine, includes multiple sites, and is undertaken on behalf of someone other than the person doing the search'';  53\% of health seekers report that most recent health information session had some kind of impact on how they take care of themselves or care for someone else.

The revolution in health care caused by the new information society is also prognosed by Haux et al. \cite{Haux_IJMI_2002} in several dimensions. Two of their theses are ``Patients and their families will be knowledgeable of the information resources available over the Internet and will make use of them. New services will arise'' (they predict the number of access to health web sites will increase by more than 30\% and that over 20\% of patients will inform themselves about health on the Internet) and that ``Knowledge about diseases will be current, comprehensive and internationally available via electronic media, including their availability to patients and their family members (`consumers'). This knowledge will be available in different qualities. Therefore, internationally accredited certification will be available for their contents (e.g. by specialty associations). Knowledge support will partially be integrated in clinical routines.'' (they predict that over 80\% of polled medical knowledge will result from electronic media and that over 80\% of the guidelines used routinely in clinical work, will be available electronically).





\section {Health Information}
\label{HI}
Information is a concept hard to define and may be viewed in several ways. It's frequent to see its definition through its comparison to the concepts of data and knowledge. Data consist of the observations or measurements made about the world. Information is the aggregated and organized data that describes a specific situation. Knowledge is what is learned from data and information, accumulated and integrated over time, that can be applied in new situations.

Current times are integrated in the third era of technological evolution, the so-called information era \cite{Webster_book_2002}. Never before, so much information has been created and transfered. A 2003 study estimated that new stored information grew about 30\% a year between 1999 and 2002 \cite{Lyman_url_2003}. In fact, information is in the center of our society and it has become an indispensable resource to every society's areas.

In healthcare, information plays a crucial role in professionals activities and consumers attitudes. Two old studies (one from 1966 and other from 1973) predicted that healthcare personal spend about one-third of their time handling and using information \cite{Jydstrup_HSR_1966, Mamlin_Medcare_1973}. According to Hersh \cite{Hersh_book_2003}, it is likely that the time dedicated to managing information in healthcare is as large, or even larger, nowadays.

In this report, the term Health Information is adopted instead of Biomedical Information, because it's broader than the latest, including not only concepts from biological and medical sciences but also information on related areas such as health care (e.g.: health care facilities, manpower and services; health care economics and organizations).

\subsection{Classification}

Textual Health Information may be classified in patient-specific information and knowledge-based information \cite{Hersh_book_2003, Shortliffe_book_2003}. The first type relates to individual patients and its purpose is to tell health professionals about the health condition of a patient. It typically comprises the patient's medical record and it may contain unstructured data (e.g.: lab results, vital signs) or free/narrative text (e.g.: radiology report). The second type of classification is related to the information that derives and is organized from observational and experimental research. This information provides health professionals the knowledge acquired in other situations so it may be applied to individual patients or used to conduct further research. Similarly to other types of scientific information, the knowledge-based information can be subdivided in primary information (direct results of original research that appears in papers, journals or other sources) and secondary information (reviews, condensations, summaries of primary literature like books, monographs, review papers, clinical guidelines, health information on web pages and other sources).

Another way of classifying knowledge-based information is to divide it in four subcategories: bibliographic, full-text, databases/collections and aggregations  \cite{Hersh_book_2003, Shortliffe_book_2003}. Each subcategory will be described next along with some of its main examples.

\subsection{Bibliographic content}

The bibliographic content contains literature reference databases (or bibliographic databases), web catalogs and specialized registries. The distinction between these subcategories is becoming blurry as, for example, literature reference databases started providing links to the referenced literature, moving closer to web catalogs.

\textbf{Literature Reference Databases}
These databases catalog books and periodicals and were the original IR databases in the past 1960s, designed to guide the searcher and not to provide information. 

MEDLINE (Medical Literature Analysis and Retrieval System Online) is probably the best-known and premier bibliographic database of the National Library of Medicine (NLM) \cite{medline_url_und}. It started being the print publication of \emph{Index Medicus}, a print catalog of the medicine literature, with its first volume published in 1879  \cite{Hersh_book_2003}. It contains over 16 million references to journal articles of approximately 5200 journals with a subject scope of biomedicine and health. In 2007, 670,000 references were added.

MEDLINE is freely available via PubMed\footnote{http://pubmed.gov} and a search generates a list of citations (including authors, title, source, and often an abstract) to journal articles, an indication of free electronic full-text availability (generally through PubMed Central\footnote{http://www.pubmedcentral.nih.gov}) or a link to the website of the publisher or other full text provider. It may also be searched using the NLM Gateway, a single Web interface that searches multiple NLM retrieval systems \cite{medlineFactSheet_url_und}. Other websites also provide access to MEDLINE, some for free and others for some fee (usually providing value-added services).

Besides MEDLINE, NLM has many other databases and electronic resources (listed in \cite{nlmDatabases_url_und}). Their bibliographic databases are organized in three categories: citations to journals and other periodicals since 1966 (accessible through PubMed that is composed by MEDLINE, MEDLINE in-process citations and publisher-supplied citations), citations to books, journals and audiovisual material (available through LOCATORplus\footnote{http://locatorplus.gov}) and citations to journal articles prior to 1966 and scientific meeting abstracts (accessible through NLM Gateway\footnote{http://gateway.nlm.nih.gov}).

In addition to NLM, there are other producers of bibliographic databases, both public and private. Some are produced by other USA National Institute of Health's institutes like the National Cancer Institute (NCI). Some of these bibliographic databases tend to be more focused in specific resource and subject types like CINAHL (Cumulative Index to Nursing and Allied Health Literature\footnote{http://www.cinahl.com}) -- the major non-NLM database for the nursing field.

\textbf{Web Catalogs}
Web catalogs are web pages that contain links to other pages and sites and share many features with traditional bibliographic databases \cite{Hersh_book_2003}. The number of such catalogs is increasing \cite{Shortliffe_book_2003}. Some well-known catalogs are: MedicalMatrix\footnote{http://www.medmatrix.org}, Hardin.MD\footnote{http://www.lib.uiowa.edu/hardin/md}, HealthFinder\footnote{http://www.healthfinder.gov}, HON Select\footnote{http://www.hon.ch/HONselect}, Intute: Health and Life Sciences\footnote{http://www.intute.ac.uk/healthandlifesciences} and MedWeb Plus\footnote{http://www.medwebplus.com}.

\textbf{Specialized Registries}
Specialized registries differ from literature reference databases because they point to more diverse information resources. This type of information resource may overlap with literature reference databases and web catalogs but, generally, it points to more diverse information resources. One famous specialized registry is the \emph{National Guidelines Clearinghouse} (NGC\footnote{http://www.guideline.gov/}), produced by the Agency for Healthcare Research and Quality (AHRQ),  with information about clinical practice guidelines.

\subsection{Full-text content}

This subcategory contains online versions of the complete version of periodicals, books and websites. Originally, full-text databases were mainly online versions of journals and it only started to include books with the decrease of the price of computers and the growth of the Internet that also led to development of web sites. 

Most periodicals are nowadays published electronically. Some are published by the company that produces the print version of the journal (e.g.: Elsevier), others by Highwire Press (which provides a website, searching and browsing  interfaces and other tools to publishers that haven't moved directly to the online world), some have created their own site and others have exclusively published online. Electronic versions are usually enhanced by additional features such as: easier access, provision of additional data like figures, tables, raw data and images, true bibliographic links. Nowadays, there are publishers that have decided to make their papers freely available on the Web. Some highly visible approaches are BiomedCentral\footnote{http://www.biomedcentral.com} (BMC), Public Library of Science\footnote{http://www.plos.org} (PLOS) and PubMed Central\footnote{http://pubmedcentral.gov} (PMC). 

Textbooks are also increasingly publishing versions on the Web. These electronic versions allow several additional features over the printed versions: they feature high-quality images; the insertion of multimedia content (e.g.: sound and video) which is impossible in the printed version; the insertion of links to other resources; the inclusion of interactive self-assessment questions and an easier access to book updates. There are already several online versions of well-known books in the health area.

The third type of full-text content is composed of web sites that provide full-text information. This excludes web sites that implement services such as bibliographic databases, online versions of books and other printed material, specialized databases and collections and aggregations of these. The Web has several health full-text information sites that are developed by everyone, from consumers to governments. Some of these websites, that include more than just collections of text (e.g.: interaction with experts, links to other sites) are: Intelihealth\footnote{http://www.intelihealth.com/}, Netwellness\footnote{http://www.netwellness.com/}, WebMD\footnote{http://www.webmd.com/}, eMedicine\footnote{http://www.emedicine.com/}, Medscape\footnote{http://www.medscape.com/} and Institute for Clinical Systems Improvement guidelines\footnote{\url{http://www.icsi.org/guidelines_and_more/}}. The first three sites are directed to consumers while the last three are more oriented to health professionals.

\subsection{Databases and Collections}

This category consists of databases and other specific collections of information. This kind of information is usually housed in database management systems and contains several types of resources like images (from radiology, pathology and other areas), genomics (gene sequencing, protein characterization and others), citations (which link scientific literature) and Evidence Based Medicine (EBM) resources. The dynamic nature of web databases make them more appropriate to some type of content. 

Images, an important piece of the healthcare practice, education and research are one of those types. There are several health images databases available on the Web. One of the most famous is the Visible Human Project\footnote{\url{http://www.nlm.nih.gov/research/visible/visible_human.html}}, which consists of three-dimensional representations of normal male and female bodies built from cross-sectional slices of cadavers.  
 
Genomics studies the genetic material in living organisms and its research has been evolving rapidly in the last years. One of its main driving forces was the Human Genome Project, led by the National Human Genome Research Institute at the National Institutes of Health (NIH). This project ended in April 2003 with the production of a version of the human genome sequence that is freely available in public databases \cite{hgp_url_und}. Several genomics databases are available across the Web and at the center are those produced by the National Center for Biotechnology Information (NCBI). NCBI's databases are linked among themselves, along with PubMed in the NCBI's Entrez system\footnote{http://www.ncbi.nlm.nih.gov/Entrez/}.

The number of citations to a specific scientific work is usually used as an evaluation measure of that work's quality and impact. To track citations in scientific literature there are citation databases that provide linkages to articles that cite others. The best-known citation databases are the  Science Citation Index (SCI) and the Social Sciences Citation Index (SSCI) from Thomson Reuters, available through the Web of Science service\footnote{http://scientific.thomson.com/products/wos/}. 

On the Web there are also databases that hold the principles of EBM and try to eliminate the problems of scattering and fragmentation of the primary literature. These databases provide systematic reviews (e.g.: The Cochrane Database of Systematic Reviews\footnote{http://www.cochrane.org}) or high synthesized synopses of evidence-based information (e.g.: Clinical Evidence\footnote{http://www.clinicalevidence.com}, DynaMed\footnote{http://www.dynamicmedical.com}, PIER\footnote{http://pier.acponline.org}, UpToDate\footnote{http://www.uptodate.com}).

\subsection{Aggregations}
This last category includes aggregations of the first three categories for all types of users, from consumers to health professionals and scientists. The distinction between this category and some of the above content with several links is blurry but, typically, aggregations have a larger variety of information that serve diverse needs of their users. They are, for example, websites that collect several types of content to generate a coherent resource. 

One of the largest aggregated consumer information resource is MedlinePlus, a service of the U.S. NLM and the NIH that is updated daily. It aggregates information from these entities and from other trusted sources on over 750 diseases and conditions. It also has preformulated MEDLINE searches to give access to medical journal articles, information on drugs, an illustrated encyclopedia, a medical dictionary, links to clinical trials, interactive patient tutorials and updated health news \cite{medlinePlus_url_und}. CancerNet is another consumer-oriented service from the National Cancer Institute\footnote{http://www.cancer.gov} (NCI) that contains information on all disease's aspects. Healthline\footnote{http://www.healthline.com} is another service focused on the consumers with online health search, content and navigation features. Kosmix RightHealth\footnote{http://www.righthealth.com} creates a page for health related topics based on information (in several formats) dispersed across the Internet. The two last consumer aggregations have been recently created and combine this service with others. The first is Microsoft HealthVault\footnote{http://www.healthvault.com}, a hub of a network of sites, personal health devices and other services to let consumers manage their health. The second is Google Health that has the same goal of giving control to health consumers, storing health information in one central place. Besides allowing the creation of online health profiles, the import of medical records, the searching of doctors and hospitals and the connection to online health services, it also aggregates resources about health topics and issues.

There are also aggregated content more directed to professionals like: MDConsult\footnote{http://www.mdconsult.com} -- a service of Elsevier that aggregates medical resources in an integrated way to help health professionals and Merck Medicus\footnote{http://www.merckmedicus.com} -- developed my Merck, available to all licensed US physicians, which includes resources like Harrison's Online and MDConsult.

\section{Health Concept Representation}
\label{HCR}

\subsection{Introduction}
The area of IR greatly benefits from the availability of well defined information structures that can be used in the indexing and retrieval processes of IR. Health information is, by its nature, highly detailed \cite{Shortliffe_book_2003}, where knowledge organization is one of the oldest applications of classification, dating to Aristotle's effort in biology and formal descriptions \cite{Pellegrin_book_1986}. Health concepts' representation is more challenging than in many domains, due to its levels of precision, complexity, implicit knowledge and breadth of application \cite{Shortliffe_book_2003}. However, it is also an area where great efforts have been developed and several representation's systems have appeared.

Before describing some of the main health concept representation systems, they are classified in three categories with increasing degree of formalism: terminologies, controlled vocabularies/thesaurus and ontologies. A terminology or vocabulary is a list of terms, that are representations of the concepts used in a specific area. When simple relationships among different terms are specified, this representation system becomes a controlled vocabulary or a thesaurus. The relationships are typically of three types:  hierarchical (terms are broader or narrower), synonymous, related (terms with relations that are neither hierarchical or synonymous). Ontologies are the most formal category having also logical descriptions that serve to computationally define terms. They must also exhibit internal consistency, acyclic polyhierarchies and computable semantics \cite{Shortliffe_book_2003}. Recently there has been an explosion of modern biomedical ontologies \cite{smith_medinfo_2004}.

The Health Concept Representation Systems described in this section may be used in several health informatics' research areas such as Information Retrieval, Natural Language Processing, Semantic Interoperability and Decision Support Systems. In Information Retrieval, they may be used in the indexing process (manual indexing is usually done using thesaurus) and in the retrieval process (e.g.: synonymous relations may be used to improve the expression of the information needs).

This section will start to describe the NLM's thesaurus that is used to index most of the NLM's databases, followed by other non-NLM thesaurus used in the health area. Then, NLM's Unified Medical Language System, together with its 3 knowledge sources: Metathesaurus, Semantic Network and the SPECIALIST Lexicon and Tools are presented. In the end, some of the main health ontologies and two general ontologies are briefly presented.

\subsection{Medical Subject Headings}
The Medical Subject Headings (MeSH) is the NLM's thesaurus used to index most of the NLM's databases \cite{Coletti_JAMIA_2001}. It has sets of terms naming descriptors that are arranged in both alphabetic and hierarchical structure (which allows searching at various levels of specificity). The 2008 version of MeSH has 24,767 descriptors organized in an eleven-level hierarchy of headings \cite{mesh_url_und}. the MeSH vocabulary files may be downloaded from the NLM site, at no charge, in XML or ASCII format.

The hierarchies in which descriptors are placed are also called trees\footnote{2008's MeSH list of trees in: http://www.nlm.nih.gov/mesh/trees2008.html}. Each Descriptor appears in at least one place in the trees and may appear in as many additional places as may be appropriate. XML MeSH is structured in three levels: Descriptor/Concept/Term. A Descriptor may consist of a class of Concepts, that, correspond to a class of Terms which are synonymous with each other. For example\footnote{Obtained from:  \url{http://www.nlm.nih.gov/mesh/concept_structure.html}}:
\begin{itemize}
\item{Cardiomegaly [Descriptor]}
\begin{itemize}
\item{Cardiomegaly [Concept, Preferred]}
\begin{itemize}         
\item{Cardiomegaly [Term, Preferred]}
\item{Enlarged Heart [Term]}
\item{Heart Enlargement  [Term]}
\end{itemize}
\end{itemize}
\begin{itemize}
\item{Cardiac Hypertrophy [Concept, Narrower]}
\begin{itemize}
\item{Cardiac Hypertrophy [Term, Preferred]}
\item{Heart Hypertrophy [Term]}
\end{itemize}
\end{itemize}
\end{itemize}

Each Concept has a Preferred Term which is also the name of the Concept and each Descriptor has a preferred Concept. The name of the Descriptor corresponds to the preferred Term of the preferred  Concept.  The terms in one Concept are not strictly synonymous with terms in another Concept, even in the same record. Additionally, MeSH has two types of relationships: hierarchical and associative \cite{Nelson_NLM_2001}. The first type is a crucial component of a thesaurus and is formalized by the MeSH tree structure that represents distinct levels of specificity (terms that are broader or narrower). MeSH descriptors are organized in 16 categories \cite{meshtree_url_und} that can be explored through the browser available at \footnote{http://www.nlm.nih.gov/mesh/MBrowser.html}.  The associative relationships are often represented by the ``see related" cross reference. They can be used to add/suggest terms to a specific search, to point out in thesaurus the existence of other descriptors which may be more appropriate or to point out distinctions made in the thesaurus or in the hierarchical structure of the thesaurus.

Besides the existence of descriptors (also called MeSH headings), MeSH has additional types of vocabulary: qualifiers (or subheadings), check tags, publication characteristics and supplementary concept records. Qualifiers can be attached to descriptors  to narrow the focus of a term (e.g.: drug therapy, diagnosis, etiology, surgery). For example, a deficiency of monoamine oxidase is retrieved by the Descriptor Monoamine Oxidase combined with the Qualifier deficiency \cite{meshRetrieval_url_und}. There are rules restricting the attachment of certain Qualifier (the allowable qualifiers are mentioned in the field \emph{Allowable Qualifiers} for each term). Check tags are a special class of MeSH descriptors that must be considered routinely for every article (that's why they're called check tags) and represent facets like species, gender, human age, historical time periods and pregnancy \cite{medlineIndexing_url_und}. Publication characteristics (or types) describes the item being indexed instead of its topic. It has 3 main categories: publication components (e.g. English Abstract), publication formats (e.g. lectures, letter), study characteristics (e.g. clinical trial, meta-analysis). The Supplementary Concept Records 
allow the indexing with headings from other thesaurus (non-MeSH headings) \cite{medlineIndexing_url_und}. Currently, MeSH has 172,000 headings in the Supplementary Concept Records within a separate thesaurus \cite{mesh_url_und}.

MeSH's use is not restricted to indexing NLM's databases. It is also used by other organizations to index bibliographic content, such as health libraries and the National Guidelines Clearinghouse\footnote{http://www.guideline.gov} \cite{Hersh_book_2003}.
          

\subsection{Non-NLM thesaurus}

In addition to Mesh there are other thesaurus in the health area used to index documents.

CINAHL\footnote{http://www.cinahl.com}, a database of nursing and allied health literature, uses the CINAHL Subject Headings, which is based on MeSH and has additional domain-specific terms \cite{Brenner_BMLA_1989}.
EMBASE\footnote{http://www.embase.com}, an european database of biomedical and pharmacological information, has a vocabulary called EMTREE\footnote{http://www.info.embase.com/emtree/about} with features similar to those of MeSH\footnote{Comparison at: \url{http://www.info.embase.com/emtree/about/emtree_mesh_comparison.pdf}}.

Other common vocabularies in the health area include the Logical Observation Identifier Names and Codes (LOINC) that provides a universal code system for reporting laboratory and other clinical observations in electronic messages \cite{McDonald_CC_2003}, HL7 vocabulary tables\footnote{http://www.hl7.org/Special/committees/Vocab/vocab.htm} that identify, organize and maintain coded vocabulary terms used in HL7 messages and the National Drug Code Directory\footnote{http://www.fda.gov/cder/ndc}.

A last vocabulary worth notice is the Consumer Health Vocabulary\footnote{http://www.consumerhealthvocab.org}, developed in an open source and collaborative initiative, that aims to help bridge the communication gap between consumers and health professionals. With this vocabulary, technical terms like ``myocardial infarction" may be translated into lay language like ``heart attack". In practice it can be used to improve IR systems, to help lay-people read and understand health-related information and others. This vocabulary has an online browser\footnote{http://samwise1.partners.org/vocab} and is organized by concepts and its associated terms.

\subsection{Unified Medical Language System}

The Unified Medical Language System (UMLS) started at the NLM, in 1986, by the hands of its Director, Donald Lindberg, as a ``long-term research and development project" \cite{Lindberg_MIM_1993}. This project aimed at reducing barriers to the application of computers to the health area and more specifically to the effective retrieval of machine-readable information \cite{Lindberg_MIM_1993, Humphreys_JAMIA_1998}. Two of such barriers are the variety of ways to express a same concept in different vocabularies and the diffusion of useful information among different systems. In fact, the medical informatics field is characterized by a large diversity of vocabularies developed for specific applications  (e.g.: epidemiological systems, medical expert systems, indexing literature, codes for billing and procedures and others). The lack of a common language barred the interoperability of the applications that used these vocabularies and was a motivation to the development of the UMLS.

The UMLS consists of three knowledge-sources that can be used separately or together. One is the Metathesaurus that has more than one million biomedical concepts from over 100 sources (including MeSH), another is the Semantic Network with 135 broad categories and 54 relationships between categories, the last one is the SPECIALIST Lexicon and Tools which has lexical information and programs for language processing \cite{Kleinsorge_NLM_2007}. The uses of these knowledge-sources can be very diverse (e.g. information retrieval, natural language processing, automated indexing, thesaurus construction, electronic health records and others). Each knowledge-source will be described in more detail in this section.

The UMLS is made available, at no cost, as 3 separate sets of relational files that are distributed on DVD or can be downloaded from the UMLS Knowledge Source Server (\footnote{http://umlsks.nlm.nih.gov/}) and 3 tools: the MetamorphoSys (UMLS installation wizard and customization tool included in each UMLS release -- \footnote{http://www.nlm.nih.gov/research/umls/meta6.html}), the  RRF Subset Browser (to find a term within a customized Metathesaurus subset or any vocabulary in the Rich Release Format (RRF) format) and the LVG (set of tools and data that are distributed with the UMLS as part of the SPECIALIST system). 


\textbf{Metathesaurus} The UMLS Metathesaurus is a multi-source (thesauri, classifications, code sets, and lists of controlled terms used in patient care, health services billing, public health statistics, indexing and cataloging biomedical literature, and/or basic, clinical, and health services research \cite{umlsMetaFactSheet_url_und}), multi-purpose and multi-lingual thesaurus of health related concepts, their various terms and the relationships among them. It is called a Metathesaurus as it transcends the thesauri, vocabularies and classifications it covers \cite{umlsMetaFactSheet_url_und}.

The list of the source vocabularies included in the 2008AA UMLS's release is available in \url{http://www.nlm.nih.gov/research/umls/metaa1.html}. The Metathesaurus isn't a new, single and unified standard vocabulary \cite{Humphreys_HPML_1993} and it doesn't contain logical assertions across terms from different vocabularies. Rather, it establishes conceptual linkages between its source vocabularies preserving their meanings, concept names and relationships. 

In the Metathesaurus, synonymous terms are clustered into a concept with a unique identifier (CUI). 
Each term, identified by a unique identifier (LUI), is a normalized name and may have several strings (identified by SUI), which represent the terms' lexical variants in the source vocabularies. Each string is associated with one or more atoms (identified by AUI) that represent the concept name in the source. For example (from \cite{Kleinsorge_NLM_2007}), the concept Headache (C0018681) has the following structure:
\begin{itemize}
\item headache (L0018681)
\begin{itemize}
\item headaches (S1459113)
\begin{itemize}
\item headaches (A1412439) -- BI
\end{itemize}
\item Headache (S0046854)
\begin{itemize}
\item Headache (A2882187) -- SNOMED
\item Headache (A0066000) -- MeSH
\end{itemize}
\end{itemize}
\item cranial pain (L1406212)
\begin{itemize}
\item Cranial Pain (S1680378)
\begin{itemize}
\item Cranial Pain (A1641293) -- MeSH
\end{itemize}
\end{itemize}
\item cephalgia head pain (L0290366)
\begin{itemize}
\item HEAD PAIN CEPHALGIA (S0375902)
\begin{itemize}
\item HEAD PAIN CEPHALGIA (A0418053) -- DxP
\end{itemize}
\end{itemize}
\end{itemize}

If two different terms have different meanings (e.g. cold) they are assigned the same LUI that stays associated with different CUI (e.g. cold temperature, common cold, cold sensation). The same can happen with strings and concepts. 

The Metathesaurus is distributed in two formats: Original Release Format (ORF) and Rich Release Format (RRF). The first is centred on the concept level. The second appeared later and is a source-centric view that supports source transparency (it has the atom level that represents the original source information and also new data fields and files to capture source specific identifiers and mappings). The access to the Metathesaurus can be made through the UMLSKS or the RRF Browser in MetamorphoSys.
 
\textbf{Semantic Network} The Semantic Network is an upper-level ontology in the health field \cite{Chen_book_2005} composed of 135 \emph{Semantic Types}, which may be assigned to Metathesaurus' concepts and 54 \emph{Semantic Relationships}, a set of relations that may hold between Semantic Types. The Semantic Types are the nodes in the Network and the Relationships are the links. It is provided in a relational table format and in a unit record format \cite{umlssemanticnetwork_url_und}.

Semantic Types are organized into 2 hierarchies: \emph{Entity} and \emph{Event} and its current scope is very broad, allowing the semantic categorization of a wide range of terminology  \cite{umlssemanticnetwork_url_und}. Each Metathesaurus's concept is assigned at least one semantic type (the most specific type available in the hierarchy). Instead of adding semantic types to the Network to encompass an object in the most appropriate categories, concepts that don't belong in a granularity level, must be associated with a Type of an upper level. For example, the Semantic Type \emph{Manufactured Object} has two child nodes: \emph{Medical Device} and \emph{Research Device}. If an object is neither a medical device nor research devices, it is simply assigned to the more general type \emph{Manufactured Object} \cite{umlssemanticnetwork_url_und}.

Semantic Relationships may be hierarchical or associative (non-hierarquical). The \emph{isa} link is the primary link in the Network and is the one that establishes the hierarchy of types and relations (e.g.: animal \emph{isa} organism and treats \emph{isa} affects). The set of associative are grouped into five major categories (which are also relations): \emph{physically related to}, \emph{spatially related to}, \emph{temporally related to}, \emph{functionally related to} and \emph{conceptually related to}. Whenever possible, relations are defined between the highest level semantic types and, generally, are inherited by all the children of those types. The relations do not necessarily apply to all instances of concepts that have been assigned to the semantic types that  are the nodes of that link. If it makes no sense, the inheritance of relations may also be blocked to a single or all of the children of the semantic types that is links . For example, ``\emph{conceptual part of} links \emph{Body System} and \emph{Fully Formed Anatomical Structure}, but it should not link \emph{Body System} to all the children of \emph{Fully Formed Anatomical Structure}, such as \emph{Cell} or \emph{Tissue}" \cite{umlssemanticnetwork_url_und}. A portion of the UMLS Semantic Network can be seen in \footnote{\url{http://www.nlm.nih.gov/research/umls/META3_Figure_3.html}}.

\textbf{SPECIALIST Lexicon and Tools} The SPECIALIST Lexicon and Tools has two main components: the lexicon and the tools. The Lexicon is a general English lexicon of common words that includes many biomedical terms and was developed to provide information to the SPECIALIST Natural Language Processing (NLP) System \cite{umlslexical_url_und}. The Tools are programs that process terms. Together they pre-process terms before their introduction in the Metathesaurus and they are very useful to NLP applications in health domain.

The SPECIALIST Lexicon's entries record the syntactic (how the words are put together), morphological (inflection, derivation and compounding) and orthographic (spelling) information of each term/word. Lexical items may be composed by more than one term, if it is how it is presented in English, medical dictionaries or thesauri, or if it is an expansion of generally used acronyms and abbreviations.

Each unit lexical record has attributes (which are called \emph{slots}) and values (which are called \emph{fillers}) and is delimited by braces (\{\}). The unit lexical records for ``anaesthetic" (one as a noun entry and the other as an adjective entry) will help to show some of the features of the Lexicon and the characteristics of lexical records (example from \cite{umlslexical_url_und}). The noun lexical record for  ``anaesthetic" is:
\begin{verbatim}
{
	base=anesthetic 
	spelling_variant=anaesthetic
	entry=E0330018
	cat=noun
	variants=reg
	variants=uncount
}
\end{verbatim}

Every lexical record has the slot \emph{base} that indicates the base form of the term. Optionally, it may also has one or more spelling variants expressed by slots \emph{spelling\_variant}. The next slot (\emph{entry}) contains the unique identifier (EUI) of the record that has an ``E" followed by seven digit numbers. The slot \emph{cat} exists in every record and indicates the syntactic category of the entry (e. g.: noun, adjective, verb). The \emph{variants} slot indicates the inflectional morphology of the entry. In the example given, these slots indicate that ``anesthetic" is a count noun which undergoes regular English formation (``anaesthetics"). In the adjective lexical record:
\begin{verbatim}
{
base=anesthetic
spelling_variant=anaesthetic
entry=E0330019
cat=adj
variants=inv
position=attrib(3)
position=pred stative
}
\end{verbatim}

\noindent the \emph{variants} slot has the filler \emph{inv} that indicates that the adjective ``anesthetic" doesn't form a comparative or superlative. The first \emph{position} slot indicates that the adjective is attributive and appears after color adjectives in the normal adjective order. The second \emph{position} slot indicates this adjective can appear in predicate position.

Other slots indicate the complementation (e. g.: in verbs if it is a intransitive or transitive verb), derivation (e. g.: adjective/noun -- red/redness) and spelling variants (e. g.: British-American variants -- centre/center) of each entry. For more detailed information see \cite{umlslexical_url_und}.

Lexical entries are independent of semantics, representing only a spelling-category pair. If different senses have the same spellings and syntactic category, they are represented by a single lexical entry in the Lexicon.

The Lexical Tools is a set of three programs implemented in Java: a lexical variant generator (lvg), a  a word index generator (Wordind) and a normalizer (NORM). They are designed to help dealing with the high degree of variability in natural language words (e. g.: treat, treats, treated, treating) and even in the order of words in ``multi-word" terms.

The lvg program performs lexical transformations of input words. It consists of several flow components that can be combined (for example, the flow \texttt{i} simply generates inflectional variants and the flow \texttt{l:i} generates the same inflectional variants but in lowercase).

The WordInd breaks strings into words and produces the Metathesaurus word index (MRXW). The use of this program before searching in the word index assures congruence the words to be looked up and the word index. The program outputs one line for each word found in the input string (e.g.: for the input string \emph{Heart Disease, Acute}, are returned three lines for the three words: \emph{heart}, \emph{disease} and \emph{acute}). The output words are always presented in lowercase.

The NORM program generates normalized strings that are used in the normalized string index (MRXNS). This program must be used before MRXNS can be accessed and it is a selection of LVG transformations (in fact, it is the lvg program with the \texttt{N} pre-selected flow option). The normalization process involves stripping possessives (e.g.: \emph{Hodgkin's diseases, NOS} -- \emph{Hodgkin diseases, NOS}), removing stop words (e.g.: \emph{Hodgkin diseases, NOS} -- \emph{Hodgkin diseases,}), lower-casing each words (e.g.: \emph{Hodgkin diseases,} -- \emph{hodgkin diseases,}), replacing punctuation with spaces (e.g.: \emph{hodgkin diseases,} -- \emph{hodgkin diseases}), breaking a string into its constituent words/uninflecting (e.g.: \emph{hodgkin diseases} -- \emph{hodgkin disease}) and sorting the words in alphabetic order (e.g.: \emph{hodgkin disease} -- \emph{disease hodgkin}).

For more information on the SPECIALIST lexicon, lexical variant programs, and lexical databases, see \cite{McCray_CAMC_1994}.



\subsection{Ontologies}

An initial definition of ontology referred to the set of primitive entities that describes and models a specific knowledge domain  and should reflect its underlying reality \cite{Gruber_und_2007}. In computer science, ontology means an organization of concepts in domains, exhibiting internal consistency, acyclic polyhierarchies and computable semantics \cite{Chen_book_2005}.

A health ontology aims to study classes of health significant entities such as substances (e.g.: mitral valve), qualities (e.g.: diameter of the left ventricle) and processes (e.g.: blood circulation) \cite{Shortliffe_book_2003}. 
In theory an ontology is different from a terminology because it is concerned with the definition of health classes and the relations among them, while the last just collects names of health entities. In practice, this distinction is less obvious and often the products developed fall in between terminologies and ontologies as they are more than lists of terms but do not necessarily meet the requirements of formal organization \cite{Shortliffe_book_2003}.

Ontologies are increasingly playing an important role in medical informatics research \cite{Musen_InfMed_2002} (e.g.: natural language processing, interoperability, SemanticWeb) where they also act as an enabling resource in several applications. Some of the main health ontologies and two other general ontologies will be presented next.

GALEN (Generalised Architecture for Languages, Encyclopedias, and nomenclatures in medicine) was the name of an European Union project (1992-1999) that illustrated how medical concepts could be represented as a formal ontology and how this could be used in practical applications \cite{Rector_CMPB_1993}. One of this project's core features is the Common Reference Model, an ontology that aims to represent ``all and only sensible medical concepts", whose access is made through OpenGALEN\footnote{http://www.opengalen.org}. GALEN provides the blocks required for describing terminologies and a mechanism for combining concepts. For example, it has explicit representations for \emph{adenocyte} and for \emph{thyroid gland} and instead of having one for \emph{adenocyte of thyroid gland}, it has an indication that these concepts can be combined. 
GALEN has an hierarchies of categories and a ``rich hierarchy of associative relationships used to define complex structures" \cite{Shortliffe_book_2003}.

As said before, when describing the Semantic Network component of the UMLS, this component serves as a basic, high-level ontology for the biomedical domain \cite{McCray_CFG_2003}. Its semantic types are used to categorize UMLS concepts that, in turn, must be assigned to, at least, one semantic type.

The Systematized Nomenclature of Medicine Clinical Terms (SNOMED CT) is the most complete biomedical terminology developed in native description logic formalism \cite{Shortliffe_book_2003}. It has a good concept coverage with over 361,800 concepts (as of July 2004), organized in eighteen independent hierarchies \cite{Melton_JBI_2006}. Each concept is described by several characteristics such as unique identifiers, parent(s), name(s) and role(s) or semantic relation(s). It is freely available as part of the UMLS.

Cyc\footnote{http://www.cyc.com/} is a general ontology that is built around a core of more than 1 million hand-coded assertions that capture ``common sense" knowledge. Groups of assertions that share a common set of assumptions (e.g.: domain, level of detail, time interval) are called \emph{microtheories}. One of these \emph{microtheories} is OpenCyc, the upper level and publicly available part of the Cyc with 6,000 concepts and 60,000 assertions on these concepts \cite{Shortliffe_book_2003}.

WordNet \cite{Fellbaum_book_1998} is an electronic lexical database used by applications of natural language processing and information retrieval. It is a general English resource composed of set of synonyms (synset) with different structures for nouns, verbs, adjectives and adverbs. This resource may be used in conjunction with other concept representations like UMLS, that don't include all word-level synonyms and permutations in health concepts, providing another component to medical concept representation and retrieval. The latest version of WordNet (3.0) contains 82,115 noun synsets categorized into hierarchies that are not based on a formal ontology theory. For example, \emph{vasoconstriction} is not formally related to the health domain. Instead it is related to \emph{constriction} what emphasizes the physical mechanism rather than pathology. WordNet is also freely and publicly available for download.



In \cite{Shortliffe_book_2003}, the representation of blood in the several systems is analysed, showing the differences among them and the richness of ontologies when compared to taxonomies. Blood is a complex case because it has two different superordinates: tissue and body substance. This is highlighted by the differing representations that raise compatibilities' issues among ontologies. The richness of ontologies is also emphasized by the additional knowledge about blood through concept's properties and through the associative relations to other concepts (in taxonomies the arrangement of concepts is only made in is-a hierarchies).

\section{Health Specific Search Engines}
\label{HSSE}
This section presents some health specific search engines with special characteristics that may be useful in this domain. Three of them are still in a beta version (Healia, MEDgle and Medstory).

Healia\footnote{http://www.healia.com} is a health search engine directed to consumers developed over more than four years with an award from the National Cancer Institute. According to their creators, its main differences to other search engines are: high quality search results (through a quality index they've developed to the analysis of health content's quality), availability of ``Personal Search" filters that allows the finding of more personally relevant information, use of health thesaurus to guide search, availability of two interfaces (a simpler one without filters and a standard one with filters).
 
MEDgle\footnote{http://www.medgle.com} is a MEDical GLobal Electronic computer generated search to general medical conditions through symptoms, diagnoses, physicians, drugs and medical procedures. In addition to the traditional text boxes of search engines, its interface is also graphic and has several filter options. Its results present not only external links but also a brief summarization of web contents about the searched topic.

Fealth\footnote{http://www.fealth.com} is the \emph{Advanced Search} product of Findica, an information and technology enterprise that develops products and services to Health Care Companies.

Medstory\footnote{http://www.medstory.com} is a Microsoft health search engine directed to consumers (Health tab) and health professionals (Research tab). It uses information from other users' searches to help the user refine and guide his search, delivering more than the standard list of web pages. It offers high-level health categorizations (e.g.: Drugs \& Substances, Conditions, Procedures), different among consumers and professionals, that allow the refinement of searches. In addition, it offers another type of high-level categorizations, according to the type of resource returned (e.g.: Web, News/Media, Audio/Video, Clinical Trials, Research Articles).

Textmed\footnote{http://www.textmed.com} is a search engine to find information about several types of medical entities  (e.g.: disease, drug, chemical, organism). It currently has 88,545 entities obtained from the analysis of about 15 million Medline abstracts. In each results' page it presents the juxtapositions (co-occurence terms) for the searched term and its relationships to other entities. Finally it presents the three top articles with that entity.


\section {Recent Work}
\label {RW}

Latest times have been of great developments in HIR. This can be confirmed by the attention given by major IR related companies like Google and Microsoft, by the number of recent created search engines (a large number of them is still in beta version) and by the large number of publications. Globally, research has been dedicated to find more efficient ways to adjust queries to user needs, to index health data, to present search results, to rank results (e.g. based on readability, content) and also to analyze information needs and behaviours.


This section describes the most recent research directions in HIR, organized by high-level research topics. 


\subsection {Health Information Seeking}

With the increase of the availability of health information on the Web and the engagement of users in health information seeking in the Web, studies of information needs and behaviors in HIR proliferate. Usually, research in this area involves not only informatics but also areas like information science, psychology and medicine. These studies produce results useful to more technical studies that aim to improve all the retrieval process, to contribute to the understanding of the Internet's influence on health behaviors and, ultimately, in the health care system. Some research papers focus on the consumer, some on the health professionals and others on global aspects of the retrieval process, done either by consumers or by health professionals. 

\subsubsection{Consumers}
In the consumer arena, a large number of studies are dedicated to characterize consumers' health information needs and their behavior. These studies are usually based on the application of surveys and log analysis. The Pew Internet \& American Life Project\footnote{http://www.pewinternet.org}, that explores the impact of the Internet in several aspects of society, has already published two reports that characterize online health search: one of the year 2000 \cite{fox_Pew_2000} and other of the year 2006 \cite{Fox_und_2006}. Rains \cite{Rains_JHC_2007} discuss the perceptions of the Web's use to seek health information gathered from the Health Information National Trends Survey's findings. These studies are centered in the USA citizen and there are others that characterize the behaviors of others countries' citizens. Two recent examples focus on portuguese \cite{Santana_AMP_2007} and greek citizens \cite{Halkias_TeH_2008}. 

There are also studies that analyze the use patterns of specific health information retrieval systems like one that, through the analysis of MEDLINEplus logs, studies the stability of user queries overtime \cite{Scott-Wright_AMIA_2006}. Others are dedicated to explore the user's information search process (e.g.: explore student's search process and outcome in Medline to write an essay for an EBM class \cite{Huuskonen_JD_2008}) and to investigate the influence of technical knowledge and cognitive abilities on health information seeking \cite{Sharit_TOCHI_2008}. Yoo's paper \cite{Yoo_JASIST_2008} has a more theoretical approach, attempting to explain how and why middle-aged women use health web sites based on two theories (the uses and gratifications approach from mass communication research and the theory of planned behavior from social psychology). An earlier and more comprehensive paper is the one from Cline \cite{Cline_HER_2001} that covers aspects like: potential benefits of health information seeking on the Internet, information quality problems, criteria for evaluating online health information and design features problems (like disorganization and technical language). 

\subsubsection{Professionals}
There are also studies that seek to understand the information needs of health professionals. Revere \cite{Revere_BI_2007} does a literature review on these needs, trying to answer the following questions: ``What are the information needs of public health professionals?", ``In what ways are those needs being met?", ``What are the barriers to meeting those needs?" and ``What is the role of the Internet in meeting information needs?". A González-González's paper \cite{Gonzalez-Gonzalez_AFM_2007}, analyzes spannish primary care physicians' information needs that arise in office practice and their information seeking patterns to satisfy these needs (in and after the consultations). Twose has a paper \cite{Twose_HILJ_2008} with the same goals but a different approach that included the combination of usage statistics from a web portal (that allowed the access to a library's electronic resources), self-report and observational data collected during an offered course. Other papers are more dedicated to the study of information seeking behaviors, such as \cite{Hemminger_JASIST_2007} where Hemminger, through a census survey, analysed the information seeking behavior of academic scientists of basic science and medical science departments; and one that determines how good is Google to lead doctors to a correct diagnosis \cite{Tang_BMJ_2006}.

Still directed to professionals, other lines of research include the study of: information retrieval systems' impact on professionals' performance in answering clinical questions (in \cite{westbrook_JAMIA_2005} this is done with a pre/post intervention experimental design and several types of health professionals), the assessment of the effectiveness of information retrieval systems for professionals (in \cite{Hersh_JAMA_1998} it is defined a framework that is used to assess Medline articles of evaluation studies), changes in information behavior after certain changes like the introduction of a clinical librarian service \cite{Urquhart_JMLI_2007} and collaborative information seeking behavior in the context of medical care \cite{Reddy_IPM_2007}.

\subsubsection{Consumers and Professionals}
There are also general studies that are neither directed to consumers or professionals. Some papers are more theoretical, globally studying health information seeking behavior (\cite{Lambert_QHS_2007} examines scientific literature from 1982 to 2006 on this subject to capture the concept's characteristics and discuss operationalizations, antecedents, and outcomes), others explore certain aspects in health information seeking behavior like user navigation \cite{Graham_AMIA_2006}, education disparities \cite{Lorence_CB_2007} and changes in user needs \cite{adams_JCDL_2005}.

Some studies are more focused on web aspects. Spink's \cite{Spink_IPM_2004} report findings from an analysis of health queries to different web search engines, providing insights into health querying and suggesting implications of the use of web search engines for health information seeking. Another paper \cite{Eysenbach_AMIA_2003} studies the prevalence of health-related searches on the Web, based on the proportion of pages on the web cointaining the search string and the word \emph{health}.

\subsection{Indexing}

Indexing in HIR is a research area than can largely benefit from the diversity and quantity of Health Concept Representations (see Section \ref{HCR}). In fact, a significant number of papers about this topic use at least one of these representations. In a Douyere's paper \cite{Douyere_HILJ_2004} the MeSH thesaurus is adapted to the broader field of health Internet resources and is used together with the Dublin Core metadata format to catalogue and index French health resources. In \cite{hliaoutakis_HIKM_2006}, MeSH is also combined with a well-established method for extraction of domain terms in the development of an automatic term extraction method for indexing large medical collections such as MEDLINE. Houston et al. \cite{Houston_DSS_2000} explore the use of concept spaces (automatically generated thesauri, where concepts are represented as nodes and relationships as weighted link, with associative memory that allows new paradigms for knowledge discovery and document searching) in HIR. They evaluated and compared the use of terms suggested by MeSH, UMLS MetaThesaurus and the automatic generated thesauri with document collection's terms. No statistically significant differences among the thesauri were fund and there was almost no overlap of relevant terms suggested by different thesauri what suggests that recall could be significantly improved using a combined thesaurus approach.

There are also papers dedicated to the development of Health Concept Representations. Zeng is a researcher with a large work on this field, being the coordinator of the Consumer Health Vocabulary Initiative\footnote{http://www.consumerhealthvocab.org/}. One of her papers \cite{Zeng_MEDNET_2006} describes the development of computerized methods to mark up Web content. Another paper  \cite{zou_doceng_2007} from a different author also presents research on segmenting and labeling HTML medical journal articles through a hidden markov model approach. In  \cite{kipp_JCDL_2007}  it is is suggested the use of social bookmarking (such as the tags from CiteULike) as an additional health concept representation and a way to discover documents not yet indexed in on-line databases.

\subsection {Retrieval}

\subsubsection{Health Information Characteristics}

One of the greatest concerns on HIR is related to the quality of published information quality and possible the interpretations given by non-experts. The analysis of health information quality (e.g. accuracy, timeliness, accessibility) and its adjustment to the user (e.g. readability) are popular lines of research in HIR. The evaluation of these two factors may be used in the results' ranking. Some of the studies in this area are described next.

In \cite{Berland_JAMA_2001} it was done an accessibility, quality and readability evaluation of health information on breast cancer, depression, obesity and childhood asthma available in English and Spanish. The accessibility of search engines was assessed using a structured search experiment, the content's quality was evaluated by physicians using structured implicit review and the reading grade level was assessed using an established method (Fry Readability method). They found that coverage of key information is poor but the accuracy is generally good and that high reading levels are necessary to comprehend health contents.

Yan et al. \cite{Yan_CIKM_2006} express the need of new computational models of readability to rank results in information retrieval systems as traditional readability formulas are too generalist. Facing this necessity, they propose a concept-based model of text readability that takes into account textual genres of a document and domain specific knowledge in three major readability formulas. 

Miller et al. \cite{Miller_HICSS_2007} also state that traditional readability formulas are not targeted to specific domains like health as they ignore the use of specialized vocabulary. In their paper they propose a na\"{i}ve Bayes classifier for three levels of health terminology specificity (consumer, health learner, health professional) created with the lexicon of a medical corpus. This classifier attained an accuracy of 96\% and was applied to consumer health web pages. Only 4\% of pages were classified as consumer ones, while all the others were included at the professional level. Miller was the second author of a recently published paper \cite{Leroy_JASIST_2008} that also describes the evaluation of the naïve Bayes classifier that was compared with readability formulas and the readability assessment of an expert and a consumer. The classifier indicated that documents were at a lower level of readability difficulty than the readability formulas. A previous paper from Leroy and other authors \cite{Leroy_AMIA_2006} compared four types of documents: easy and difficult WebMD documents, patient blogs, and patient educational material. They found that it is possible to simplify many documents based on terminology in addition to sentence structure (but this can still be insufficient for difficult documents).

In \cite{Keselman_ISBMDA_2006} the males and females' familiarity with terms of three types of health topics (male-specific, female-specific and gender-neutral) is evaluated. It was found that males were more familiar with neutral and male-specific topics and that females has no topic effect. In face of these results, the tailoring of health readability formulas to target populations is also discussed. Rosemblat et al. \cite{Rosemblat_MEDNET_2006} have done an exploratory study to analyze the relevance of readability's predictors in the consumer health domain, based on expert judgment and to characterize expert ratings' patterns across the various predictors. 
They concluded that the development of health readability tools may require the modification of existing measures (e.g. including health-related vocabulary) and the addition of new predictive features.

Zeng is the first author of three papers related to health consumer terminology. The first \cite{Zeng_MIM_2002} studied the characteristics of consumer terminology used in HIR through the log analysis of two consumer web sites and patients' interviews. They concluded that there are significant mismatches between consumer and information source terminologies. In \cite{Zeng_ISBMDA_2005} it was created a method to measure the familiarity of medical terms and a predictive model for familiarity, based on term occurrence in text corpora and reader's demographics. In the third paper \cite{Zeng_AMIA_2005} the authors developed a systematic methodology using text analysis and human review to assign \emph{consumer-friendly} names to UMLS's concepts. The evaluation of this method was done applying a questionnaire to consumers and the results suggested this methodology is useful in the development of consumer health vocabularies.

\subsubsection{Query Expansion}

Is often happens that  query terms are related to terms used to index documents but are not indexing terms. This motivates the development of techniques of query expansion. These methods are used to improve precision in search results with alternative/additional query terms (synonyms or other semantic relationships). This technique is quite used, probably due to the large number of Health Concept Representations available.  

In another Zeng's paper \cite{zeng_JAMIA_2006}, a tool to assist people in health-related query construction was developed. The suggested terms were selected based on their semantic distance to the original query (calculated through co-occurences in medical literature and log data and also through semantic relations in medical vocabularies). They concluded that semantic-distance-based query recommendations can help consumers with query formulation during HIR.

In \cite{Fattahi_IPM_2008} it is proposed a query expansion method through text analysis of non-topical terms in Web documents. They define the concepts of topical terms (TT) as the terms that represent the subject content of documents (e.g.: breast cancer), non-topical terms (NTT) are, usually, terms that occur before or after topical terms to represent a specific aspect of the subject (e.g. `about' in `about breast cancer') and semi-topical terms (STT) that are terms that normally do not occur alone, being used in conjunction with topical terms to narrow or further specify the subject -- they are normally domain-specific (e.g.: `risk of' in `risk of breast cancer'). The defined method to query expansion is based on the use of NTT and STT in conjunction with TT. 

Ide et al. \cite{Ide_JAMIA_2007} describe the algorithms used in a search engine with query expansion and probabilistic relevancy ranking, evaluated using data and standard evaluation methods from the 2003 and 2006 TREC Genomics Track.

\subsubsection{Ranking}

The ranking algorithms used in a IR system, responsible of adjusting the position of each result in the returned results' list, can also be used to enhance these systems (more specifically, the precision at first returned results). Price is a co-author with, at least, two publications about approaches to improve ranking in HIR. In \cite{Price_AMIA_1999} the authors present a system that ranks results according to the likely quality of page health contents. In \cite{Price_AMIA_2005} the approach to improve the results' ranking is different. The authors model queries as relationships between concepts and try to match these relations with the ones existing between documents. The analysis of users' browsing behavior is also a typical approach to the development of ranking algorithms (e.g.: \cite{Anagnostopoulos_ISU_2007}) .

\subsubsection{IR Models}

The definition of IR models is another topic seen in HIR papers. The use of semantic information is typical in the most recent proposed models. This is is the case of the model proposed by Price et al. in \cite{price_HIKM_2006} and \cite{Price_CIKM_2007}. In this model, the authors describe the content of documents in domain-specific collections using semantic components (``segments of text about a particular aspect of the main topic of the document that may not correspond to structural elements in the document" \cite{Price_CIKM_2007}), complementary to full text and keyword indexing. In the first paper \cite{price_HIKM_2006}, the authors introduce the model, present the results of its application to the representation of clinical questions in the medical domain and present ways to use the model for retrieval. In the second paper \cite{Price_CIKM_2007}, the authors present experimental evidence that the model enhances the retrieval of domain-specific documents in response to real users' realistic queries.

In \cite{Hliaoutakis_ECDL_2006} an IR model, implemented in MedSearch (a Medline retrieval system), that discovers similarities between documents containing semantically similar but not necessarily lexically similar terms is presented. In \cite{Dung_DaWak_2007} it is presented a methodology to build and enhance an ontology in health domain through semantic elements extraction. The extracted information from Web documents is then summarized, indexed and stored in the database for an implemented information retrieval system.

A very recent paper \cite{Hung_JBI_2008} presents an information seeking model to represent human search expertise that may allow the development of an intelligent search agent that generates adaptive search strategies based on the human search expertise. The model described is hierarchical and multi-level where each level represents a problem space traversed during the search process and a layer of knowledge required to a successful search.

\subsection {Evaluation}




The evaluation of an IR system isn't a simple task, involving frequently human intervention (in the judgement of relevance or in users' studies). Another approach involves the use test collections (set of documents and questions) and less human intervention. An example are the ones used in TREC\footnote{http://trec.nist.gov}. The development of such collections is the target of some papers. For example, in \cite {hersh_jamia_2006}, Hersh et al. have developed a test collection to assess visual and textual methods in biomedical image retrieval. 
 
An example of a user centered study is the one presented in \cite{kushniruk_AMIA_2002} where a comparative usability evaluation of an automated text summarization system and three search engines is done. The evaluation involved audio and video recording of subject interactions with the interfaces. Another paper in which evaluation is done with human intervention is one from Tang et al. \cite{Tang_IR_2006}. In this paper, human assessors were used to do the relevance judging according to a scheme preciously developed. The goal was to compare the performance of domain-specific health and depression search engines with Google on both relevance of results and quality of advice.

\subsection {User Interfaces and Visualization}

The development of user interfaces and aspects of information visualization in IR systems is another area with a significant quantity of research work. In the health area, with developments in concept representation models, it also became popular the use of this information to improve the way results are presented to the user.

In \cite{Stuckenschmidt_IEEEIS_2004} is presented a system that implements a concept-based visualization of the results that, according to a user study, is less suitable for searching specific information and more suitable to the exploration of mostly unknown data. In \cite{Douyere_HILJ_2004}, a paper already cited in this report, a terminology is developed from the MeSH thesaurus and metadata elements and is used in several tasks, one of them being the visualization and navigation through the concept hierarchies. Stapley et al. \cite{stapley_PSB_2000} have built a system for retrieving and visualizing co-occurences of gene names in Medline abstracts. From the co-occurence data is built a graph where nodes are genes and edge lenghts are a function of the co-occurence of the two genes in the literature. 






\section{People}
\label {People}





The most notorious person in HIR is William Hersh\footnote{http://medir.ohsu.edu/~hersh/}, M.D. and Professor and Chair of the Department of Medical Informatics \& Clinical Epidemiology in the School of Medicine at Oregon 
Health \& Science University. His research focuses on the development and evaluation of IR systems for health practitioners and researchers. He has a several publications in this field, is the author of a book with an Information Retrieval's health perspective \cite{Hersh_book_2003}, the author of Health Informatics'  book chapters (like \cite{Shortliffe_book_2003, Chen_book_2005}) and the editor of other Health Informatics' books (like \cite{Chen_book_2005}).

Another person with a significant research activity in HIR is Qing Treitler Zeng\footnote{http://dsg.harvard.edu/~qzeng/}, an Assistant Professor of the Decision Systems Group in Harvard Medical School and a Research Associate at the Brigham and Women's Hospital. One of her research interests is the semantic knowledge-based information retrieval and presentation. One of her latest research projects, in which she assumes the role of leader, is the Consumer health vocabulary\footnote{http://www.consumerhealthvocab.org/} which aims to build, in an open source and collaborative way, a vocabulary that links everyday words about health with technical terms used by professionals. One of this project's applications is in the area of Information Retrieval as stated in \cite{zeng_JAMIA_2006}, one of her many publications in the HIR field.

Susan L. Price (no personal webpage was found) is thought to be a PhD Student at the Department of Computer Science of the Portland State University, supervised by Professor Lois Delcambre. She has been working on the use of semantic components to enhance retrieval of domain-specific information (being health the domain selected as a case study) and is the author of papers with Lois Delcambre, Marianne Lykke Nielsen of the Royal School of Library and Information Science in Denmark and William Hersh.

Other authors with some publications in the area are David Howard Hickman\footnote{\url{http://www1.va.gov/pshsrd/docs/CV_Hickam_02.htm} and \url{http://www.bio-computing.org/showauthor.php?surname=Hickam&initials=DH}} with jointed works with William Hersh; Angelos Hliaoutakis\footnote{http://www.softnet.tuc.gr/~angelos/}, a MSc student in Computer Engineering that has been publishing in the HIR area; Thanh Tin Tang\footnote{\url{http://www.informatik.uni-trier.de/~ley/db/indices/a-tree/t/Tang:Thanh_Tin.html} and \url{http://libra.msra.cn/authordetail.aspx?id=1316585&query=web+search}} and Madhu C. Reddy\footnote{http://faculty.ist.psu.edu/reddy/Research.htm} which has interest in Collaborative information behavior and has been developing research work in the area of healthcare.

There are also authors with a less straight connection to the specific area of HIR and more connected to the broader fields of eHealth and Health Informatics. However, having a straight connection to the health area and published work in HIR, they will be mentioned next. Gunther Eysenbach\footnote{\url{http://gunther-eysenbach.blogspot.com/}} is an Associate Professor at the Department of Health Policy, Management and Evaluation at the University of Toronto and one of the editors of \cite{Shortliffe_book_2003}. Susannah Fox\footnote{\url{http://www.pewinternet.org/PPF/a/104/about_staffer.asp}} is a Pew Internet \& American Life Project associate director connected to the Internet's impact on health care with several published reports\footnote{http://www.pewinternet.org/PPF/c/5/topics.asp}. Steve Rains\footnote{\url{http://datamonster.sbs.arizona.edu/communication/faculty/each_detail.php?option=1&detail=57&mtitle=Core\%20Faculty}} has also some published papers in the area of Health Information Seeking Behavior. Edward H. Shortliffe\footnote{http://www.dbmi.columbia.edu/shortliffe} is a well known name in the broad area of health informatics with specific interests in decision-support systems, integrated workstations for clinicians and web-based information dissemination. 

Three last authors worth mentioning, more connected to the development of ontologies in the health area, are James J. Cimino\footnote{http://www.dbmi.columbia.edu/~ciminoj/} that is one of the editors of \cite{Shortliffe_book_2003}, Mark Musen\footnote{\url{http://med.stanford.edu/profiles/Mark_Musen/}} and Alan Rector\footnote{http://www.cs.manchester.ac.uk/research/publications/byauthor/Rector/}.

\section{Research Groups}
\label {RG}





Only two research groups on the specific field of HIR were found and the second one is related to the fields of health and IR but is not directly related to HIR. The first is the Centre for Health Information Management Research\footnote{http://www.shef.ac.uk/chimr/research/}, at the University of Sheffield, more connected to the study of Health Information Needs and Behaviors. The second one is the W3C Semantic Web Health Care and Life Sciences Interest Group\footnote{http://www.w3.org/2001/sw/hcls/} which aims to develop, promote and support the use of Semantic Web technologies in health care and life sciences.

\section{Research Projects}
\label {RP}

The two most active persons in the area have four research projects mentioned next. William Hersh has three projects: the TREC Genomics Track retrieval of scientific literature on genomics\footnote{http://ir.ohsu.edu/genomics/}, ImageCLEFmed\footnote{http://ir.ohsu.edu/image/} that is part of the Cross Language Evaluation Forum\footnote{\url{http://www.clef-campaign.org/}} (CLEF) for medical image retrieval and OHSUMED\footnote{http://ir.ohsu.edu/ohsumed/}, a test collection of a subset MEDLINE references, created to assist in information retrieval research.

The fourth project is leaded by Qing Zeng and is entitled Consumer Health Vocabularies\footnote{http://www.consumerhealthvocab.org/}, a project dedicated to the development of vocabularies that link everyday words about health to professional terms used by health care professionals.




\section{Main Organizations}
\label{MO}

Some of the main organizations in HIR, some more connected to the health area than others, are listed below.

\begin{itemize}
\item American Library Association-- \url{http://www.ala.org}
\item American Medical Informatics Association -- \url{http://www.amia.org}
\item American Society for Information Science and Technology -- \url{http://www.asis.org}
\item American Society for Indexing -- \url{http://www.asindexing.org}
\item Association for Computer Machinery, Special Interest Group on Information Retrieval -- \url{http://www.sigir.org}
\item Medical Library Association -- \url{http://www.mlanet.org}
\item National Library of Medicine -- \url{http://www.nlm.nih.gov}
\item Special Libraries Association -- \url{http://www.sla.org}
\end{itemize}

\section{Main Journals}
\label{MJ}



Some of the main journals where HIR's papers can be published are listed below. Some of them are more related to the Computer Science field, some to the Information Science field and others to the health and health informatics field. 

\begin{itemize}
\item ACM Transactions on Information Systems -- \url{http://www.acm.org/tois}
\item Artificial Intelligence in Medicine -- \url{http://www.elsevier.com/wps/find/journal}\\ \url{description.cws_home/505627/description#description}
\item British Medical Journal -- \url{http://www.bmj.com}
\item Computers in Biology and Medicine -- \url{http://www.elsevier.com/locate/inca/351}
\item Health Information and Libraries Journal -- \url{http://www.blackwellpublishing.com/journal.asp?ref=1471-1834}
\item Journal of Biomedical Informatics -- \url{http://www.academicpress.com/jbi}
\item Journal of the American Medical Informatics Association -- \url{http://www.jamia.org}
\item Journal of the American Society for Information Science and Technology -- \url{http://www.asis.org/jasist.html}
\item Journal of the Medical Library Association-- \url{http://www.mlanet.org/publications/jmla/}
\item Informatics for Health and Social Care -- \url{http://www.tandf.co.uk/journals/tf/14639238.html}
\item Information Processing and Management -- \url{http://www.elsevier.com/locate/info}\\ \url{proman}
\item Information Retrieval -- \url{http://www.springer.com/computer/database+management+\%26+information+retrieval/journal/10791}
\item International Journal of Medical Informatics -- \url{http://www.elsevier.com/wps/find/journaldescription.cws_home/506040/description#description}
\item Medical \& Biological Engineering \& Computing -- \url{http://www.springer.com/engine}\\\url{ering/biomedical+eng/journal/11517}
\item Medical Decision Making -- \url{http://mdm.sagepub.com/}
\item Methods of Information in Medicine -- \url{http://www.schattauer.de/index.php?id=704}
\end{itemize}

\bibliographystyle{plain}
\bibliography{bibliography}
\end{document}